\documentclass{article}
\usepackage{spconf,amsmath,graphicx,bm,cite}
\usepackage{algorithm}
\usepackage{algorithmic}
\usepackage{booktabs}
\usepackage{booktabs}
\usepackage{braket}
\usepackage{breqn}
\usepackage[absolute,showboxes]{textpos}

\TPMargin{5pt}

\newcommand{\copyrightstatement}{
    \begin{textblock}{0.84}(0.08,0.93)    
         \noindent
         \footnotesize
         \copyright  2019 IEEE.  Personal use of this material is permitted.  Permission from IEEE must be obtained for all other uses, in any current or future media, including reprinting/republishing this material for advertising or promotional purposes, creating new collective works, for resale or redistribution to servers or lists, or reuse of any copyrighted component of this work in other works.
    \end{textblock}
}

\DeclareMathOperator*{\argmin}{arg\,min}
\title{Multi-channel Time-Varying Covariance Matrix Model  \\ for Late Reverberation Reduction}
\name{Masahito Togami$^1$}
\address{
  $^1$LINE Corporation, Tokyo, Japan
  }
\begin{document}
\ninept
\maketitle
\begin{abstract}
In this paper, a multi-channel time-varying covariance matrix model for  late reverberation reduction is proposed. 
Reflecting that variance  of the late reverberation is time-varying 
and it depends on past speech source variance, 
the proposed model is defined as convolution of a speech source variance 
with a multi-channel time-invariant covariance matrix of late reverberation.
The multi-channel time-invariant covariance matrix  can be interpreted as 
a covariance matrix of  a multi-channel acoustic transfer function (ATF).  
An advantageous point of the covariance matrix model  against a deterministic ATF model  
is that the covariance matrix model is robust against fluctuation of the ATF.  
We propose two covariance matrix models. 
The first model  is a covariance matrix model of late reverberation in  the original microphone input signal.
The second one is a covariance matrix model of late reverberation  
in an extended microphone input signal which includes 
 not only  current microphone input signal but also past microphone input signal. The second one 
 considers correlation between the current microphone input signal and the past microphone input signal.
Experimental results show that the proposed method effectively reduces reverberation especially 
in  a time-varying ATF scenario and the second model 
is shown to be more effective than the first model.

\end{abstract}
\begin{keywords}
Speech dereverberation, covariance matrix model, non-negative convolutive transfer function, time-varying acoustic 
transfer function
\end{keywords}

\section{Introduction}
\label{sec:intro}

\copyrightstatement

 
Speech dereverberation techniques which reduce harmful room reverberation 
have been actively studied  \cite{naylor} for speech quality improvement of human-listening devices 
and for automatic speech recognition performance improvement. 
When multiple microphones are available, theoretically, complete dereverberation can be achieved 
by multi-channel spatial inverse filtering under the assumption that acoustic transfer functions (ATFs)
 are given in advance \cite{MINT}. 
Two stage methods which consist of a blind channel identification \cite{huang2005} 
and a multi-channel spatial inverse filtering have been commonly utilized. 
Simultaneous estimation of the ATFs and  an anechoic speech source 
have been also studied \cite{NIPS2000_1908,KEMDEREVERB,Schmid2014,schwartz2015}. 
 However, estimation of the ATFs is typically unstable.

Recently, auto-regressive (AR) model
based speech dereverberation techniques  have been actively studied 
\cite{Nakatani2010,TASLPTOGAMI2013,braun2016,dietzen2017,Otsuka,Kagami,Giacobello2018}. 
Instead of the ATF estimation, the AR coefficients of the microphone input signal are estimated.
The AR coefficients can be easily calculated by estimating only correlation of the microphone 
input signal along the time axis.  Estimation of the AR coefficients can be performed more stably than estimation of ATFs.
However, the AR model based methods assume that the ATF is time-invariant. 
Therefore, when the ATF is time-varying, dereverberation performance degrades.

So as to reduce reverberation when the ATF is time-varying, dereverberation methods with 
probabilistic ATF models have been also studied \cite{Schmid2014,Laufer2019,togami_interspeech2019}.
When both the speech sources and the ATF are probabilistic variables, parameter optimization based on the 
expectation-maximization (EM) algorithm \cite{em} is not applicable, because probabilistic models 
for joint latent variables are hard to be defined. Instead of the EM algorithm, parameter estimation methods with variational Bayesian approximation
 have been studied \cite{Schmid2014, togami_interspeech2019}.
 In the probabilistic ATF model, not only the mean vector of the ATF but also the covariance matrix 
 of the ATF are utilized for speech dereverberation based on a time-varying multi-channel Wiener filtering. 
Thus, it is shown that a covariance matrix term is highly effective for speech dereverberation when 
the ATF is time-varying. 

In the single-channel speech dereverberation research field, another late reverberation model 
in the non-negative spectral domain has been studied, i.e., non-negative convolutive transfer function (N-CTF) model \cite{kameoka2009,singh2010,Yu2012ConstrainedMS,Mohammadiha,N-CTF}. 
The N-CTF assumes that the power spectral of the late reverberation 
is convolution of power spectral of a speech source with a non-negative transfer function.
The N-CTF model optimizes the parameters based on an auxiliary function approach. 
The dereverberated signal can be obtained by a time-varying filter.
The N-CTF model is robust against the time-varying ATF, because sensitivity against 
the time-varying ATF is caused by the phase term of the ATF and the phase term is neglected in the N-CTF model.
However, the N-CTF model does not utilize spatial information effectively even when a multi-channel microphone input 
signal is available.
 
Inspired by the effectiveness of a multi-channel covariance matrix model in the  speech dereverberation context, 
we propose a multi-channel covariance matrix model of late reverberation in this paper so as to reduce 
reverberation especially when the ATF is time-varying. 
The proposed method models a multi-channel covariance matrix of late reverberation 
more directly than the conventional probabilistic ATF models.
Reflecting that variance  of the late reverberation is time-varying 
and it depends on past speech source variance, 
the proposed model is defined as convolution of the speech source variance 
with a multi-channel covariance matrix of late reverberation.
The speech source variance is modeled by the N-CTF. Thus, the proposed multi-channel 
covariance matrix model of the late reverberation is a natural extension of the 
N-CTF model into a multi-channel model with spatial information. 
We propose two covariance matrix models. 
The first model  is a covariance matrix model of the late reverberation in  current
microphone input signal. 
The second model is a covariance matrix model of the late reverberation 
in an extended microphone input signal which includes 
 not only  current microphone input signal but also past microphone input signal. 
The second model considers a cross-correlation between the current microphone input signal 
and the past microphone input signal. 
All of the parameters in the proposed method can be updated similarly to 
 the multi-channel non-negative matrix factorization (MNMF) \cite{SAWADA2013,icml_yoshii2013}
  so as to decrease a cost function monotonically
  with an auxiliary-function approach.

\section{Problem statement}
\subsection{Signal model}
In this paper, dereverberation is performed in a time-frequency domain.
Multi-channel microphone input signal $\bm{x}_{l,k}$ in the time-frequency domain 
($l$ is the frame index and $k$ is the frequency index) 
is modeled as follows:
\begin{equation}
\bm{x}_{l,k}=\sum_{d=0}^{L_{d}-1} s_{l-d,k} \bm{a}_{l,d,k} + \bm{v}_{l,k}, \label{input_model}
\end{equation}
where $\bm{x}_{l,k}$ 
is the multi-channel microphone input signal at each time-frequency point, 
the number of the microphones is $N_{m}$, 
$s_{l,k}$ is the speech source signal, $\bm{a}_{l,d,k}$ is the time-varying acoustic 
transfer function (ATF), $L_{d}$ is the tap-length of the ATF, and $\bm{v}_{l,k}$ is the background noise term. 
Since there is a scale ambiguity problem between $s_{l,k}$ and $\bm{a}_{l,d,k}$,
the objective of speech dereverberation is defined as extraction of $s_{l,k}\bm{a}_{l,d=0,k}$ 
from the observed microphone input signal $\bm{x}_{l,k}$.

\subsection{Non-negative convolutive transfer function model}
Non-negative convolutive transfer function (N-CTF) model \cite{N-CTF} reduces reverberation with the following approximated microphone input signal model:
\begin{equation}
|x_{l,k,m}|^{p} \approx \sum_{d=0}^{L_{d}-1} |s_{l-d,k}|^{p}|a_{l,d,k,m}|^{p}, \label{nctf_model}
\end{equation}
where $p$ is set to $1$ or $2$. $x_{l,k,m}$ is the $m$-th element of $\bm{x}_{l,k}$ and 
$a_{l,d,k,m}$ is the $m$-th element of $\bm{a}_{l,d,k}$.
The background noise term is neglected in the N-CTF model. 
Parameter optimization of the N-CTF model can be done efficiently via the non-negative matrix factorization (NMF) \cite{NMF_LEE2000}.
The N-CTF model is robust against the time-varying ATF, because the phase term is removed in Eq.~\ref{nctf_model}. 
However, the N-CTF model based on  Eq.~\ref{nctf_model} does not utilize spatial information even 
when a multi-channel microphone input signal is available.

\section{Proposed method}
\subsection{Overview}
We propose a multi-channel extension of the N-CTF model which utilizes spatial information effectively 
when a multi-channel microphone input signal is available.
The proposed method utilizes the original microphone input signal model defined in Eq.~\ref{input_model}.
The proposed method introduces the N-CTF model 
into a multi-channel covariance matrix model of  late reverberation.
We propose two multi-channel covariance matrix models, i.e., a covariance matrix model of the original microphone input signal
and a covariance matrix model of the convolutive microphone input signal. 
In both models, 
speech dereverberation is performed based on a time-varying spatial filtering with the proposed multi-channel 
covariance matrix model of the late reverberation. 
Similarly to the multi-channel non-negative matrix factorization (MNMF) \cite{SAWADA2013}, All parameters 
are updated so as to increase a likelihood function of microphone input signal monotonically by using an auxiliary function approach.



\subsection{Covariance matrix model of original microphone input signal}
In the proposed method, the likelihood function of the multi-channel microphone input signal 
is maximized w.r.t. parameters. The likelihood function of the microphone input signal is defined based on the local Gaussian modeling \cite{duong2010} as 
follows: 
\begin{equation}
p(\bm{x}_{l,k} | \theta_{k})=\mathcal{N}(0, \bm{R}_{x,l,k}),
\end{equation}
where $\bm{R}_{x,l,k}$ is the time-varying covariance matrix of the multi-channel microphone input signal 
and $\theta_{k}$ is the separation parameter. 
Multi-channel covariance matrix $\bm{R}_{x,l,k}$ can be also modeled as follows:
\begin{equation}
\bm{R}_{x,l,k}=\bm{R}_{r,l,k} + \bm{R}_{v,k},
\end{equation}
where $\bm{R}_{r,l,k}$ is the multi-channel covariance matrix of the late reverberation, 
 and $\bm{R}_{v,k}$ is the multi-channel covariance matrix of the background noise. We assume that the noise signal 
is stationary and $\bm{R}_{v,k}$ does not depend on the frame index $l$.  
Since the amount of the late reverberation reverberation depends on the variance of the past speech source, $\bm{R}_{r,l,k}$
 also depends on  the variance of the past speech source. Thus, 
$\bm{R}_{r,l,k}$ is modeled as the following convolution of the time-frequency variance of the speech source 
and the time-invariant multi-channel covariance matrix:
\begin{equation}
\bm{R}_{r,l,k}=\sum_{d=0}^{L_{d}-1} v_{l-d,k} \bm{R}_{d,k},  \label{cov_model}
\end{equation}
where $v_{l,k}$ is the time-frequency variance of the speech source and 
$\bm{R}_{d,k}$ is the multi-channel covariance matrix of the $d$-th tap of the late reverberation.
When the number of the microphone input signal is $1$, Eq.~\ref{cov_model} 
 is identical to Eq.~\ref{nctf_model} with $p=2$. Thus, the proposed covariance matrix model 
 is a multi-channel extension of the N-CTF model with the spatial model of the late reverberation $\bm{R}_{d,k}$.
 Thanks to the spatial model, late reverberation can be reduced not only by non-negative filtering but also by multi-channel spatial filtering.
 $\theta_{k}$ is defined as $\{\{v_{l,k}\}_{l}, \{\bm{R}_{d,k}\}_{d}, \bm{R}_{v,k}  \}$.

\subsection{Covariance matrix model of convolutive microphone input signal model}
Inspired by a joint integration of beamforming and speech dereverberation with a convolutive microphone input signal 
 \cite{conv_beam_nakatani2019}, we also propose a covariance matrix model for 
convolutive microphone input signal $\tilde{\bm{x}}_{l,k}$.
In the convolutive microphone input signal, not only the current microphone input signal but also the past microphone input 
signals are concatenated as follows:
\begin{equation}
\tilde{\bm{x}}_{l,k}=[\begin{array}{cccc} \bm{x}_{l,k} & \bm{x}_{l-1,k}^{T} & \cdots & \bm{x}_{l-L_{x}+1,k}^{T}  \end{array}]^{T},
\end{equation}
where $T$ is the transpose operator of a matrix/vector. 
The covariance matrix of $\tilde{\bm{x}}_{l,k}$, $\bm{R}_{\tilde{x},l,k}$, 
includes a cross-correlation between the current microphone input signal and the past microphone input signals.
Thus, speech dereverberation can be done by using the cross-correlation information. 
$\bm{R}_{\tilde{x},l,k}$ is 
modeled as follows:
\begin{equation}
\bm{R}_{\tilde{x},l,k}=\sum_{d=0}^{L_{d}-1} v_{l-d,k} \tilde{\bm{R}}_{d,k} + \tilde{\bm{R}}_{v,k}.
\end{equation}
$v_{l-d,k} \tilde{\bm{R}}_{d,k}$ is the covariance matrix term of the late reverberation 
that depends on the speech source variance at the ($l-d$)-th frame.
Considering  the independence assumption of the speech source along the time-axis and the 
causality of the reverberation, 
components of $\tilde{\bm{R}}_{d,k}$ which depend on the frames before the ($l-d$)-th frame
are set to $\bm{0}$ as follows:
\begin{equation}
\tilde{\bm{R}}_{d,k}=\begin{pmatrix}
\bar{\bm{R}}_{d,k,N_{m}(d+1) \times N_{m}(d+1)} & \bm{0} \\
\bm{0} & \bm{0} \\
\end{pmatrix}.
\end{equation}
 $\theta_{k}$ is defined as $\{\{v_{l,k}\}_{l}, \{\tilde{\bm{R}}_{d,k}\}_{d}, \tilde{\bm{R}}_{v,k}  \}$ in this case.
When $L_{d}$ is $1$, the covariance matrix model of the convolutive microphone input signal is identical to 
the covariance matrix model of the original microphone input signal. Thus, the covariance matrix model of the convolutive 
microphone input signal is a generalization of the covariance matrix model of the original microphone input signal.
From here, the covariance matrix model of the convolutive 
microphone input signal is utilized as an representative. 


\subsection{Parameter optimization}
Optimization of the parameter $\theta_{k}$ is performed based on minimization of the negative log likelihood 
function $\mathcal{F}(\theta_{k})$ that is defined as follows:
\begin{equation}
\mathcal{F}(\theta_{k}) = \sum_{l=1}^{L_{T}} \tilde{\bm{x}}_{l,k}^{H} \bm{R}_{\tilde{x},l,k}^{-1}
 \tilde{\bm{x}}_{l,k} + \log \det \bm{R}_{\tilde{x},l,k}+const.
\end{equation} 
Similarly to the multi-channel non-negative matrix factorization (MNMF) \cite{SAWADA2013,icml_yoshii2013},
an auxiliary function $\mathcal{F}^{+}(\theta_{k}, \phi_{k})$ ($\phi_{k}$ is an auxiliary variable) 
can be obtained which fulfills the following conditions:
\begin{eqnarray}
\mathcal{F}(\theta_{k}) &\le& \mathcal{F}^{+}(\theta_{k}, \phi_{k}), \\
\mathcal{F}(\theta_{k}) &=& \min_{\phi_{k}} \mathcal{F}^{+}(\theta_{k}, \phi_{k}).
\end{eqnarray}
By using an auxiliary function, the parameter $\theta_{k}$ and the auxiliary variable $\phi_{k}$ are 
updated alternately as follows: 
\begin{eqnarray}
\phi_{k}^{(t+1)} &=& \argmin_{\phi_{k} } \mathcal{F}^{+}(\theta_{k}^{(t)}, \phi_{k}), \label{phi_upate} \\
\theta_{k}^{(t+1)} &=& \argmin_{\theta_{k} } \mathcal{F}^{+}(\theta_{k}, \phi_{k}^{(t+1)}), \label{theta_update}
\end{eqnarray}
where $t$ is the iteration index. 
we can decrease the cost function monotonically based on Eq.~\ref{phi_upate} and Eq.~\ref{theta_update}, 
because 
\begin{dmath}
\mathcal{F}(\theta_{k}^{(t+1)}) \le \mathcal{F}^{+}(\theta_{k}^{(t+1)}, \phi_{k}^{(t+1)}) \le \mathcal{F}^{+}(\theta_{k}^{(t)}, \phi_{k}^{(t+1)})=\mathcal{F}(\theta_{k}^{(t)}).
\end{dmath}

An auxiliary function can be obtained as follows:
\begin{dmath}
 \mathcal{F}^{+}(\theta_{k}, \phi_{k})= \sum_{l=1}^{L_{T}} \hat{\bm{R}}_{\tilde{x},l,k} \bm{B}_{l,k}^{H} \tilde{\bm{R}}_{v,k}^{-1} \bm{B}_{l,k}
+ \sum_{l=1}^{L_{T}}  \sum_{d=0}^{L_{d}-1}  \frac{\text{tr}(\hat{\bm{R}}_{\tilde{x},l,k} \bm{Q}_{d,l,k}^{H} \tilde{\bm{R}}_{d,k}^{-1} \bm{Q}_{d,l,k})}{v_{l-d,k}}
+ \log \det \bm{U}_{l,k} + \text{tr}(\bm{R}_{\tilde{x},l,k} \bm{U}_{l,k}^{-1})-N_{m},
\end{dmath}
where $\hat{\bm{R}}_{\tilde{x},l,k}=\tilde{\bm{x}}_{l,k}\tilde{\bm{x}}_{l,k}^{H}$ and 
$\phi_{k}=\{\bm{Q}_{d,l,k},\bm{B}_{l,k},\bm{U}_{l,k}\}$. $\phi_{k}$ is updated by 
Eq.~\ref{phi_upate} as  follows:
\begin{eqnarray}
\bm{Q}_{d,l,k}^{(t+1)}&=&v_{l-d,k}^{(t)} \tilde{\bm{R}}_{d,k}^{(t)} \bm{R}_{\tilde{x},l,k}^{{(t)},-1}, \\ 
\bm{B}_{l,k}^{(t+1)}&=&\tilde{\bm{R}}_{v,k}^{(t)}\bm{R}_{\tilde{x},l,k}^{{(t)},-1},\\
\bm{U}_{l,k}^{(t+1)}&=&\bm{R}_{\tilde{x},l,k}^{(t)}. 
\end{eqnarray}
The parameter $\theta_{k}$ is updated by Eq.~\ref{theta_update} as follows:
\begin{equation}
 \tilde{\bm{R}}_{d,k}^{(t+1)}=\bm{G}_{d,k}^{(t),-1} \mathcal{\#} (\tilde{\bm{R}}_{d,k}^{(t)} \bm{J}_{d,k}^{(t)} \tilde{\bm{R}}_{d,k}^{(t)}),
\end{equation}
\begin{equation}
 \tilde{\bm{R}}_{v,k}^{(t+1)}=\bm{R}_{\tilde{x},l,k} \mathcal{\#} (\tilde{\bm{R}}_{v,k}^{(t)} \bm{E}_{k}^{(t)}  \tilde{\bm{R}}_{v,k}^{(t)}),
\end{equation}
\begin{equation}
v_{l,k}^{(t+1)} =v_{l,k}^{(t)} \sqrt{\frac{\sum_{d} \text{tr}(\bm{R}_{\tilde{x},l+d,k}^{(t),-1} 
\hat{\bm{R}}_{\tilde{x},l+d,k} \bm{R}_{\tilde{x},l+d,k}^{(t),-1}\tilde{\bm{R}}_{d,k}^{(t+1)}  )}
{\sum_{d} \text{tr}(\bm{R}_{\tilde{x},l+d,k}^{(t),-1}\tilde{\bm{R}}_{d,k}^{(t+1)})}}, \label{v_opt}
\end{equation}
where $\mathcal{\#}$ is the geometric mean operator \cite{yoshii_2018}, 
$\bm{G}_{d,k}=\sum_{l} v_{l-d,k}\bm{R}_{\tilde{x},l,k}^{-1}$, 
$\bm{J}_{d,k}=\sum_{l} v_{l-d,k} \bm{R}_{\tilde{x},l,k}^{-1} \hat{\bm{R}}_{\tilde{x},l,k} \bm{R}_{\tilde{x},l,k}^{-1}$, and 
$\bm{E}_{k}=\sum_{l} \bm{R}_{\tilde{x},l,k}^{-1} \hat{\bm{R}}_{\tilde{x},l,k} \bm{R}_{\tilde{x},l,k}^{-1}$.

After the parameter optimization, the output signal can be obtained via a multi-channel Wiener filtering as follows:
\begin{equation}
\bm{y}_{l,k}=v_{l,k} \tilde{\bm{R}}_{d=0,k} \bm{R}_{\tilde{x},l,k}^{-1} \tilde{\bm{x}}_{l,k}.
\end{equation}
We utilize the first $N_{m}$ elements of $\bm{y}_{l,k}$ as a multi-channel output signal.

\section{Experiment}
\subsection{Setup}

Speech dereverberation and noise reduction performance 
were evaluated under a time-invariant ATF scenario and a time-varying ATF scenario. 
The number of the microphones, $N_{m}$, was set to 2.
Sampling rate was set to 16000 Hz. Framesize was 1024 pt, and frame shift was 512 pt. The number of the speech sources was set to 1.
Multi-channel data was generated by convolving a measured impulse response with a clean speech source.
The clean speech sources were extracted from TIMIT test corpus \cite{timit}.
As the measured impulse responses, 
Multi-Channel Impulse Response Database \cite{MIRD} was utilized. 
Impulse responses of the 1st impulse response and the 2nd one from a linear microphone array 
with spacing ``3-3-3-8-3-3-3''(cm) were utilized.
The reverberation time $RT_{60}$ was 0.61 (sec).
The assumed tap-length $L_{d}$ was set to 6. 
The azimuth of the speech source was set to 0 degrees.
The distance between the microphones and the speech source location was set to $2$ m.
The number of the clean speech sources that utilized in the experiment was 10.
As background noise signal, we selected "office", "cafeteria", and  "meeting" noise 
from DEMAND dataset \cite{thiemann_joachim_2013_1227121}. For each noise type,  the randomly extracted noise signals were convolved with the 
impulse response from each azimuth (0,15,30,45,60,75,90,270,285,300,315,330,345 degrees) and mixed so as to mimic diffuse noise.
SNR was set to 20 dB. There were total 60 samples.
The number of the iterations was set to 20.
In the time-varying ATF scenario, the impulse response $a_{m,d=bL+l,i} $ 
($L$ was set to 4800 sample, $m$ is the microphone index, $i$ is the tap index, and $d$ is the frame index. frame was set to 256 samples)
was generated as follows:
\begin{eqnarray}
a_{m,d=bL+l,i}&=& (1-|\alpha_{bL+l}|)a_{m,\theta=0,i} \nonumber \\ 
&+& \max(0,\alpha_{bL+l}) a_{m,\theta=15,i} \nonumber \\ 
&+&\max(0,-\alpha_{bL+l}) a_{m,\theta=345 ,i},
\end{eqnarray}
where $a_{m,\theta,i}$ is the impulse response of the azimuth $\theta$ and 
\begin{eqnarray}
\alpha_{bL+l}&=& \frac{L-l}{L}\beta_{b} + \frac{l}{L}\beta_{b+1}, \\
p(\beta_{b})&\sim&\mathcal{N}(0,1).
\end{eqnarray}
For dereverberation performance evaluation, we utilized four evaluation measures which were defined in REVERB Challenge \cite{reverb_challenge2014}, 
i.e., Cepstrum distance (CD) [dB], Log likelihood ratio (LLR), Frequency-weighted segmental 
SNR (FWSegSNR) [dB], 
and Perceptual Evaluation of Speech Quality (PESQ).  

\subsection{Comparative methods}
We compared the proposed method with the following methods:
\begin{itemize}
\item WPE \cite{Nakatani2010}: We utilized implementation in \cite{Drude2018NaraWPE}
\item N-CTF: Dereverberation is performed based on the input model defined by 
Eq.~\ref{input_model}. For parameter optimization, the Itakura-Saito Divergence based loss 
function \cite {N-CTF} was utilized with $p=1,2$. 
\item Kalman-EM for dereverberation (KEMD) \cite{KEMDEREVERB}: Mixture is modeled 
as convolution of a time-invariant acoustic impulse response and a time-varying 
speech source signal.
\item Variational Bayesian  Multi-channel Speech Dereverberation (VB-MSD) \cite{togami_interspeech2019}: 
Mixture is modeled 
as convolution of a time-varying acoustic impulse response and a time-varying 
speech source signal.
\item TIV ($L_{d}=0$ and $L_{x}=6$): The extended microphone input signal model is utilized.
No time-varying covariance matrix model for late reverberation is utilized. 
The late reverberation is reduced by utilizing only a time-invariant covariance 
matrix.
\end{itemize}
As the proposed method, the following two settings were evaluated. 
\begin{itemize}
\item PROPOSED 1 ( $L_{d}=6$ and $L_{x}=0$): The original microphone input signal model is utilized.
The late  reverberation is reduced by utilizing the proposed time-varying covariance matrix model.
\item PROPOSED 2 ($L_{d}=6$ and $L_{x}=6$): The extended microphone input signal model is utilized.
The late  reverberation is reduced by utilizing the proposed time-varying covariance matrix model.
\end{itemize}

\subsection{Experimental results}
In Table \ref{evaluation_result1} and Table \ref{evaluation_result2}, 
the experimental results for the time-invariant ATF scenario and the time-varying ATF scenario 
are shown, respectively. 
It is shown that the proposed method with the 
time-varying multi-channel covariance matrix model 
is more effective
than  the other methods. From comparison between TIV and PROPOSED 1 or PROPOSED 2, 
it is shown that the proposed time-varying covariance matrix model of late reverberation 
is effective. 
From comparison between PROPOSED 1  and 
PROPOSED  2, it is also shown that the extended 
microphone input signal model is more effective than the original
microphone input signal model. In this paper, we focused on a covariance matrix model of late reverberation, but 
it is also possible to integrate the proposed covariance model with 
the mean vector similarly to \cite{TASLPTOGAMI2013}. Integration of the proposed covariance matrix model 
with the mean vector is beyond the scope of this paper, and it is one of future works.
\begin{table}[t]
\caption{Evaluation results for time-invariant ATF scenario }
\label{evaluation_result1}
\begin{center}
\begin{tabular}{ccccc}\toprule
 Approach &  CD  & LLR & FWSeg. &  PESQ \\
 &   (dB) &  & SNR (dB) &    \\\midrule
 Unprocessed  & 3.92 & 0.75 & 10.04 & 1.45 \\\midrule
 WPE & 3.52 & 0.64 & 11.50 &  1.63 \\
 N-CTF ($p=1$)   & 3.45  & 0.62  & 11.20  & 1.67   \\
 N-CTF ($p=2$)   & 3.45  & 0.62  & 11.33  & 1.64  \\
 KEMD  & 3.22 & 0.67 & 12.46 &  1.63 \\
 VB-MSD  & \textbf{3.09} & 0.62 & 12.16 &  1.76 \\
 TIV  & 3.60 & 0.68 & 11.83 &  1.54 \\ \hline
 PROPOSED 1 & 3.27 & 0.56 & 12.22 &  1.77 \\ 
 PROPOSED 2  & 3.10 & \textbf{0.52} & \textbf{12.92} &  \textbf{1.91} \\\bottomrule
\end{tabular}

\end{center}
\end{table}

\begin{table}[t]
\caption{Evaluation results for time-varying ATF scenario }
\label{evaluation_result2}
\begin{center}
\begin{tabular}{ccccc}\toprule
 Approach &  CD  & LLR & FWSeg. &  PESQ \\
 &  (dB) &  & SNR (dB) &    \\\midrule
 Unprocessed & 3.99 & 0.74 & 10.20 &  1.45 \\\midrule
 WPE  & 3.79 & 0.68 & 10.86 & 1.52 \\
 N-CTF ($p=1$)  & 3.58 & 0.61 & 11.14  & 1.65   \\
 N-CTF ($p=2$)  & 3.57  & 0.62 & 11.38  & 1.60  \\
 KEMD & 3.67 & 0.73 & 11.08 & 1.52 \\
 VB-MSD  & 3.44 & 0.67 & 11.39 &  1.64 \\
 TIV  & 3.90 & 0.74 & 10.74 &  1.42 \\ \hline
 PROPOSED 1  & 3.47  & 0.59 & 11.71 & 1.65 \\
 PROPOSED 2  & \textbf{3.40} & \textbf{0.57} & \textbf{12.36} &  \textbf{1.72} \\\bottomrule
\end{tabular}

\end{center}
\end{table}

\section{Conclusions}
In this paper, we proposed a time-varying multi-channel covariance matrix model
 for late reverberation. The proposed model is a multi-channel extension of 
 the conventional single-channel non-negative convolutive transfer function model. 
 We further proposed a time-varying multi-channel covariance matrix model
 for a convolutive microphone input signal. 
 The parameter of the proposed method can be optimized with an auxiliary function based approach. 
Experimental results showed that the proposed time-varying multi-channel covariance matrix model 
can reduce reverberation more than the conventional methods even when acoustic transfer functions are 
time-varying and that the covariance matrix model for 
a convolutive microphone input signal model was effective.
\bibliographystyle{IEEEbib}
\bibliography{mask,lgm,spbasic,gan,doa,kalman}

\end{document}